\documentclass[ showpacs, secnumarabic,amssymb,nobibnotes,aps,prl]{revtex4-2}
\usepackage{graphicx, graphics, amsmath, amssymb, rotate, textcomp, gensymb}
\usepackage{mathrsfs, float}
\usepackage[T1]{fontenc}
\usepackage{dcolumn}
\usepackage{physics}
\usepackage{wrapfig}
\usepackage{xcolor}
\usepackage[
    colorlinks=true,
    citecolor=blue
]{hyperref}

\usepackage[none]{hyphenat}
\usepackage{times}
\usepackage{multirow}
\begin{document}

\title{ Aggressive Phase Separation in Dense Mixtures of Passive and Active Particles}

\author{Purnendu Pathak}\thanks{Equal contributions}
\author{Gokul Upadhyay}\thanks{Equal contributions}
\author{Subir K. Das}
\email{Email of the corresponding author: das@jncasr.ac.in}
\affiliation{Theoretical Sciences Unit and School of Advanced Materials, Jawaharlal Nehru Centre for Advanced Scientific Research, Jakkur, Bangalore 560064, India}
\date{\today}

\begin{abstract}
Using Vicsek-like self-propulsion rule, we study kinetics of liquid-liquid phase separation in mixtures of passive and active particles.
For evolutions following temperature quenches of homogeneous configurations to the immiscible region of the phase diagram, we identify a remarkably strong dependence of the domain growth exponent on the choice of final state point. 
The singular dependence is indicative of the possibility of even an exponentially fast growth for suitable choices of system parameters.
This striking observation, supported by finite-size scaling and other advanced analyses, is despite the fact that the overall mixture density is quite high that risks congestion with the prospect of slowing down particle transport.
From flocking in pure biological systems to robotic swarming, these results are of much practical relevance. 
Theoretical pictures suitable for interpreting such high growth rates are discussed. 
In this regard, we discuss coarsening in the velocity field as well that describes how the density field coarsening follows the latter.

\end{abstract}

\maketitle

\section{Introduction}
\sloppy

Near the critical points of equilibrium phase transitions various thermodynamic properties exhibit power-law singularities \cite{Fisher1967, stanleybook, Kawasaki1972, Goldenfeld_book}.
Related exponents assume numbers that are universal for large varieties of systems and transitions. 
Though exists, universality in dynamics is much weaker \cite{ Onuki2002, Bray2002, puribook}. 
In fact, a fundamental importance of phase transition research lies in demonstrating how diverse microscopic dynamics converge into robust static universality \cite{LS1961, Fisher1967, stanleybook, Kawasaki1972, Goldenfeld_book, Hohenberg1977, Binder1974, sengers_1980, Furukawa1985, Onuki2002, Bray2002, puribook, SKD_PRL2006}.
In dynamics, phenomena and properties of interest include the critical behavior of transport coefficients \cite{sengers_1980, SKD_PRL2006, SKD_JCP2006} and kinetics following quenches of systems from homogeneous to inhomogeneous regions of various phase diagrams \cite{Bray2002, McLeish2003, puribook}.
Recently focus has shifted significantly to studying such phenomena in active matter systems \cite{vicsek95, vicsek1997, toner_tu, albano08, Gompper_rev2020, puricoarsening, Paul2020flocking, binder2021phase, paul_bera, Maity2023_twotemp, bera2022, Sengupta2025}.
Self-propelling components push these systems out of equilibrium, triggering a richer variety of transitions with superior diversity in dynamics.
Here we study phase separation kinetics in a model mixture \cite{AO_PRL2014, AO_JCP2016, Cates2015, Mishra2018, Ignacio2022, Maity2023, Manisha2025} of passive and active particles in which self-propulsion obeys a Vicsek-like velocity alignment rule \cite{vicsek95, vicsek1997, das2017, TP2024}. 

A fundamental enquiry in kinetics centers on how rapidly domains of a phase grow with time $(t)$. 
Classification of the growth phenomena \cite{Bray2002, puribook}, in passive matter, has been made on the basis of order-parameter conservation, space dimensionality, hydrodynamic effects and range of interparticle interactions.
Extraordinary variation in mechanism, as well as in growth rate, during phase separation, may also occur due to shift of final state point inside the miscibility gap, owing to change in temperature \cite{majumder2013, skd2017, midya2020} as well as that in density or composition \cite{Binder1974, Binder1977, Siggia1979, Furukawa1985, Furukawa1987, Tanaka1994, Tanaka1996, kendon2001inertial, Roy2013kinetics, majumder2013, Midyaprl, Koyel2023}.
For example, there is drop in growth rate in many passive transitions \cite{Redner2001, zerotmp_redner, blanchard2014, blanchard2017, skd2017, Chakraborty2017, Nalina2019, midya2020, nalina2022} at sufficiently low temperatures.
We note here that in the energy minimization process during evolution, interfacial tension provides a key driving force, particularly when structures are made of percolating or interconnected domains. A high value of surface tension is expected to provide strong drive for phase separation. 
Nevertheless, lowering of rate does occur \cite{Nalina2019, midya2020, nalina2022}, despite the fact that surface tension increases with the decrease of temperature \cite{Das2011PRL,Widom1992surface}, owing to formation of metastable structures \cite{Redner2001, zerotmp_redner, blanchard2014, blanchard2017, Chakraborty2017, Nalina2019}. 

Interestingly, growth exponent, in our evolving active matter system, displaying percolating structure, exhibits a sharp rise with the decrease of temperature.
This, having no passive counterpart, indicates the possibility of even exponentially rapid phase separation. 
Nevertheless, like in the standard passive matter cases, we demonstrate that growth obeys self-similarity \cite{Bray2002, puribook, cugliandolo2010topics} and finite-size scaling \cite{Fisher_barber, landaubook, Binder1996, majumder2011_2, majumder2013}.
The enhancement in the growth rate is in sharp contrast even with the results from corresponding single-component active matter systems, undergoing vapor-liquid transition, for which the growth rate remains quite insensitive to the variation of temperature. 
The exceptional rise in the rate for the liquid-liquid transition is despite the fact that overall particle density is much higher, providing the possibility of crowding. This may imply complex behavior of interfacial tension at the active-passive boundaries and enhanced transport leading to breaking of metastable barriers, if any. 
The results have important practical implications, including advantageous manipulation of collective phenomena in artificial active flocking such as in robotic swarms \cite{swarm2013, robot2014, Poschel2018, kilbot2024}. 
We discuss theoretical pictures relevant for interpretation of such fast rates of phase separation.
We also present results on velocity field which shows that ordering in the latter drives the density field phase separation.

\section{Model and Methods}

\begin{figure}
    \centering
    \includegraphics*[width=0.48\textwidth]{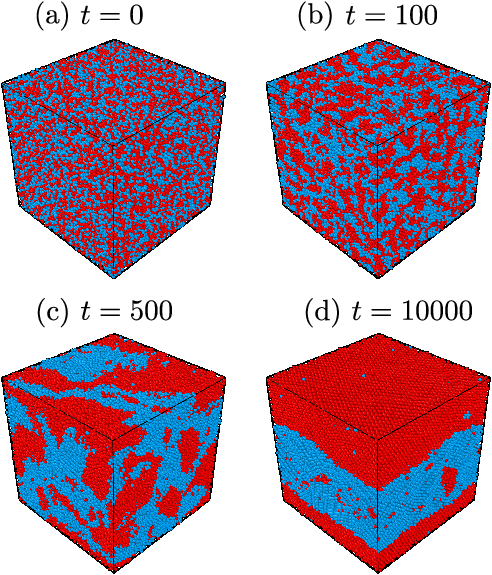}
    \caption{Demonstration of phase separation in our model mixture of active and passive particles with symmetric composition.
    For the activity strength we have chosen $f_A=5$.
    The temperature is set at $T=0.5$. 
    Presented snapshots are for a system of linear size $L=48 \sigma$.
    }
    \label{snap_fig}
\end{figure}

In our binary $(A+P)$ mixtures, for passive interparticle interactions, that prevent particle overlap, unlike the original Vicsek model, we have used a cut and shifted version of the Lennard-Jones (LJ) potential \cite{Allenbook} $(r=| \vec r_i - \vec r_j|$, the distance between $i^\text{th}$ and $j^\text{th}$ particles)
\begin{equation}
     u(r)=U(r)-U(r_c), ~ \text{for} ~ r \leq r_c.
    \label{lj_eq}
\end{equation}
Here $U(r)~(=4\varepsilon_{\alpha \beta}\left[(\sigma/r)^{12}-(\sigma/r)^{6}\right])$ is the standard LJ pair potential, with $\varepsilon_{\alpha \beta} $  ($\alpha,\beta \in [A,P]$) being the pairwise interaction strength and $\sigma$ the interaction diameter, same for all pairs. 
The cut-off distance $r_c$ is set at $2.5\sigma$ and we have chosen $\varepsilon_{AA}=\varepsilon_{PP}=2\varepsilon_{AP}=\varepsilon$. 
The latter choice provides a symmetric coexistence curve, around the critical composition $50:50$, in the pure passive limit. 
The critical temperature, $T_c$, is $1.638 ~ \varepsilon/k_B$ \cite{das2003transport}, when the overall particle density $\rho$ is set at unity, $k_B$ being the Boltzmann constant, in $d=3$, the space dimension of our interest. 
Dynamics in our work  was studied by solving the equation \cite{schlickbook, AO_PRL2014, binder2021phase,  Sengupta2025}
\begin{equation}
    m \ddot{\vec{r}}_i = -\vec{\nabla} U_i - \gamma m \dot{\vec{r}}_i + \vec{F}_i^r(t) + \vec{f}_i,
    \label{langevin_eq}
\end{equation}
the mass $(m)$ being same for all particles.
The random force $\vec{F}_i^r(t)$ is related to the friction coefficient $\gamma$, set at unity, via the fluctuation-dissipation relation \cite{schlickbook}, and satisfies $\langle F_i^{r,\mu}(t) \rangle = 0$, along with 
\begin{equation}
    \langle F_i^{r,\mu}(t) F_{i'}^{r,\nu}(t') \rangle = 2m \gamma k_B T \delta_{\mu \nu} \delta_{ii'} \delta(t - t'), 
\end{equation}
where $[\mu, \nu ] \in [ x, y, z]$. 

The self-propulsion rule for the active force, $\vec f_i ~ (=f_A \vec D_i)$, is incorporated in a way similar to the Vicsek Model (VM) \cite{vicsek95}. 
In our implemented rule, update of the velocity of a particle $i$ is influenced by the local average direction, $\vec{D}_i$, and an activity strength $f_A$, the former being computed as \cite{das2017}
\begin{equation} 
    \vec{D}_i = \frac{\sum_{j \in \mathcal{R}_c} \vec{v}_j}{\left| \sum_{j \in \mathcal{R}_c} \vec{v}_j \right|}.
\end{equation}
Here $\vec{v}_j$ is the velocity of the $j^{\rm th}$ particle lying inside a sphere $\mathcal{R}_c$ of radius $r_c$ around the $i^{\rm th}$ particle.
The active force acts in such a way that it changes only the direction of the active particles, keeping the magnitude of particle velocity, and, thus, the temperature of the system unaltered \cite{das2017, TP2024}.
We first update the velocity following the standard (passive) Verlet velocity rule \cite{schlickbook, Frankelbook}, applied to the Langevin Eq. \eqref{langevin_eq}.
This is denoted by $\vec v^{~\rm pas}$.
The final update is carried out by incorporating the direction in accordance with the activity \cite{das2017, TP2024} as
\begin{equation}
    \vec{v}_i(t+\Delta t) = \left\lvert \vec{v}_i^{~\rm pas}(t+\Delta t)\right\rvert \hat{n}; 
\end{equation}
with
\begin{equation}
    \ \hat{n} = \frac{\vec{v}_i^{~\rm pas}(t+\Delta t) + \frac{\Delta t}{m}\vec{f}_i}{\left\lvert \vec{v}_i^{~\rm pas}(t+\Delta t) + \frac{\Delta t}{m}\vec{f}_i\right\rvert}.
\end{equation}
Note that in the pure passive limit $f_A$ assumes the value zero.

We have carried out simulations in periodic cubic boxes of linear dimension $L$.
Variations were made in $f_A$, $T$ and $L$. 
For $L$ we chose values up to $64$, containing more than a quarter million particles.
For solving Eq. \eqref{langevin_eq} we have chosen $\Delta t = 0.005 \tau$, with $\tau =\sqrt{m\sigma^2/\varepsilon}$. 
Temperature, length and mass in our studies are measured in units of $\varepsilon/k_B$, $\sigma$ and $m$, respectively.
For the sake of convenience we have set $m$, $\varepsilon$, $\sigma$ and $k_B$ to unity.

\section{Results}

\begin{figure}
    \centering
    \includegraphics*[width=0.48\textwidth]{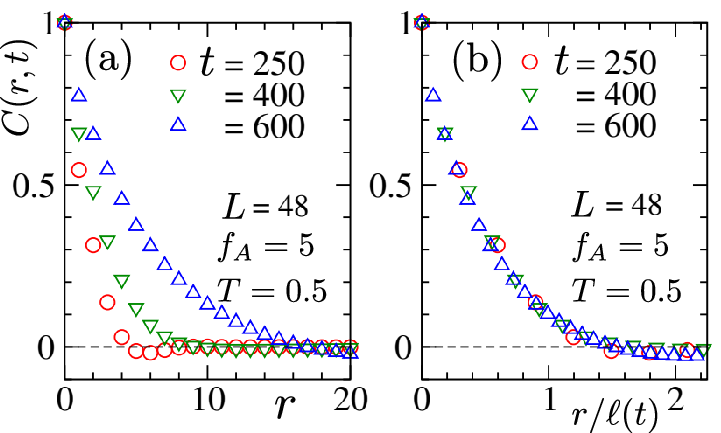}
    \caption{(a) Correlation functions, $C(r,t)$, at different times, are plotted versus $r$.
    System parameters are similar to those in Fig. \ref{snap_fig}.
    (b) Scaling plots of the same correlation functions after dividing the distance variable by time dependent length.}
    \label{cf_fig}
\end{figure}

In Fig. \ref{snap_fig} we show evolution snapshots, for the parameter combination $T=0.5$, $f_{A}=5$ and $L=48$, obtained during phase separation following quenches of homogeneous configurations to the miscibility gap.
Nice percolating structures can be identified.
Careful examination of times mentioned above the frames may suggest rapid growth.
Typically, in kinetics of phase separation the growth is self-similar in nature. 
This implies that patterns at two different times differ only by the change in average size $(\ell)$  of domains. 
This self-similarity translates to a scaling behavior of the two-point equal time $(t)$ correlation function \cite{Bray2002}:
\begin{equation}
    C(r,t) \equiv \tilde C(r/\ell),
    \label{scaling_eq}
\end{equation}
where $\tilde{C}(x)$ is a time independent master function and $C(r,t)$, for isotropic domain pattern, as applicable to the present problem, is defined as \cite{Bray2002} ($\vec r_1$, $\vec r_2$ being two different space points,  $r= |\vec{r_1}-\vec{r_2}|$)
\begin{equation}
   C(r,t) = \langle\psi(\vec{r_1},t)\psi(\vec{r_2},t)\rangle - \langle\psi(\vec{r_1},t)\rangle\langle\psi(\vec{r_2},t)\rangle.
   \label{cf_eq}
\end{equation}
In the above, $\psi(\vec r,t)$ is space and time dependent order parameter.
In Fig. \ref{cf_fig} (a) we show $C(r,t)$, as a function of $r$, for different times, with $L=48$. 
With the progress of time the decay gets slower, implying growth in the system.
To calculate $C(r,t)$, we have mapped our off-lattice systems to lattice ones \cite{majumder2011, das2017, TP2024}.
In this process $\psi(\vec r,t)$ is assigned the value $+1$ or $-1$ depending upon whether the local relative particle concentration is higher or lower than zero \cite{das2006spinodal}. 
The dynamical scaling in Eq. \eqref{scaling_eq} is demonstrated in Fig. \ref{cf_fig} (b).
Nice collapse of data from different times, upon dividing $r$ by time-dependent length, implies self-similarity, as in passive systems. 
The value of $\ell$ was obtained from the distance at which $C(r,t)$ at a given time decays to $0.1$.

\begin{figure}
    \centering
    \includegraphics*[width=0.48\textwidth]{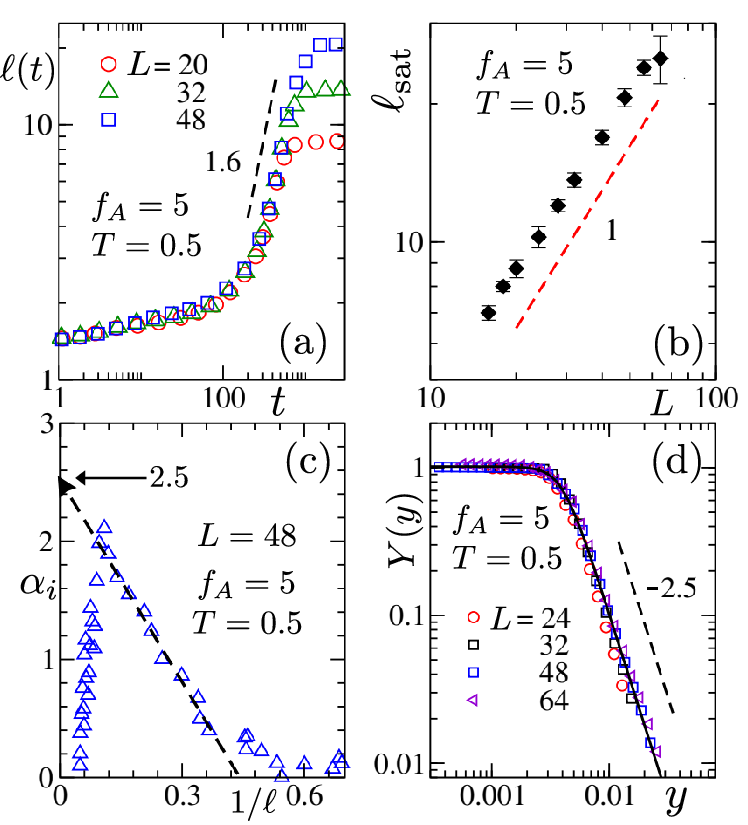}
    \caption{(a) Time evolution of the average domain length, $\ell(t)$, displayed as a function of $t$ on logarithmic axes for different system sizes. The data correspond to the same values of $f_A$ and $T$ used in Fig.~\ref{snap_fig}. The dashed line represents a power-law behaviour, with the corresponding exponent labelled alongside. 
    (b) Dependence of the saturation domain length, $\ell_{\rm sat}$, on the system size $L$. The dashed line establishes an expected linear relationship.
    (c) Instantaneous growth exponent, $\alpha_i$, plotted against $1/\ell$, for $L=48$. The arrowed dashed line serves as a visual guide to the asymptotic limit, $\ell \rightarrow \infty$. The fall in large $\ell$ implies finite-size effects.
    (d) Finite-size scaling analysis of the average domain length using the data shown in (a) for different system sizes. The continuous curve represents the analytical scaling function $Y$, while the dashed line highlights the asymptotic growth exponent.
    }
    \label{doml_Lvary_fig}
\end{figure}

In Fig. \ref{doml_Lvary_fig} (a) we show $\ell$ vs $t$ plots for $T=0.5$ and $f_A=5$. 
Data for different system sizes are included.
While there is nice agreement of data from different $L$ at early times, deviations and finally finite-size saturations occur at different time and length scales with the variation of $L$. 
In addition to the structural scaling, leading to the validity of Eq. \eqref{scaling_eq} for self-similarity, in passive matter phase transitions one typically observes finite-size scaling (FSS) \cite{Fisher_barber, landaubook, Binder1996, majumder2011_2, majumder2013}.
An important aspect of the FSS scaling is that these saturation values, as well as the deviation points, typically scale with the system size linearly.
In Fig. \ref{doml_Lvary_fig} (b) we show $\ell_{\rm sat}$, the saturation values, versus the system size.
Nice linear behavior appears, satisfying the above mentioned expectation.

Results in Fig.~\ref{doml_Lvary_fig} (a) conveys that the growth is exceptionally rapid, with $\alpha \simeq 1.6$, defined via \cite{Bray2002}
\begin{equation}
    \ell(t) \sim t^{\alpha},
\end{equation}
surpassing even the fast hydrodynamic growth, due to advective transport for percolating structure in three dimensional passive systems.
However, a log-log plot can lead to incorrect estimation of $\alpha$, if there exists off-set or a cross-over.
In such a situation other methods should be adopted.
A method, commonly used in the literature of phase transition kinetics, requires calculation of a time dependent exponent \cite{Huse1986, majumder2011_2}, viz., 
\begin{equation}
    \alpha_i = \frac{d \ln \ell}{d \ln t}.
    \label{inst_eq}
\end{equation}
In Fig. \ref{doml_Lvary_fig} (c) we plot $\alpha_i$ as a function of $1/ \ell$. 
Somewhat late time or large $\ell$ data show robust linear increase, before falling due to the appearance of finite-size effects. 
Extrapolation of the linear regime provides $\alpha \simeq 2.5$, in the thermodynamic limit, viz., for $\ell=\infty$. 
The quoted value is exceptionally high!

Another way to obtain reliable values of $\alpha$ is to carry out FSS analysis \cite{Fisher_barber, landaubook, Binder1996, majumder2011_2, majumder2013}.
Careful examination of results in Fig. \ref{doml_Lvary_fig} (c) suggests a gradual crossover, implying effects of self-propulsion become important beyond a certain critical length scale.
In such a situation it is difficult to carry out FSS with the direct data sets. 
To construct the FSS ansatz, we then write \cite{SKD_PRL2006, SKD_JCP2006, majumder2011_2}
\begin{equation}
    \ell (t) = \ell_0 + At^\alpha,
    \label{growth_eq}
\end{equation}
where $\ell_0$ serves as a background contribution that we will deal with in a manner done previously in critical phenomena \cite{SKD_PRL2006, SKD_JCP2006}. 
Unlike the critical contribution, a background contribution in the critical vicinity displays weak dependence on relevant parameter like temperature, corresponding variable in the present case being time. 
Here note that analogous to the steady-state correlation length $\xi (T) \sim \epsilon^{-\nu}$, $\epsilon \propto |T-T_c|$, in critical phenomena, one has $\ell (t) \sim (\frac{1}{t})^{-\alpha}$ in coarsening phenomena.
For the FSS purpose, the weakly varying background can be treated as constant \cite{SKD_PRL2006, SKD_JCP2006}.
For the current problem, the background $\ell _0$, should be comparable to the crossover length. 
This is acceptable given that the pre-crossover growth is weak.
From the exercise in Fig. \ref{doml_Lvary_fig} (c), $\ell_0$ appears to be $\simeq 2.25$. 


From Eq. \eqref{inst_eq} and \eqref{growth_eq} it appears that \cite{Amar1988, majumder2011_2}
\begin{equation}
    \alpha_i = \alpha \left[1-\frac{\ell_0}{\ell}\right].
    \label{inst2_eq}
\end{equation}
The simulation data are indeed consistent with the linear behavior noted in Eq. \eqref{inst2_eq}.
The latter suggests that the slope of the line in Fig. \ref{doml_Lvary_fig} (c) should be $-\alpha \ell_0$.
This indeed is true, as can be readily verified from Fig. \ref{doml_Lvary_fig} (c), when we use $\alpha=2.5$ and $\ell_0=2.25$. 
This justifies the consistency between different analyses.

Using Eq. \eqref{growth_eq} one can construct the following FSS scaling ansatz:
\begin{equation}
    \ell(t) - \ell_0 = Y(y)\left(\ell_{\rm sat} - \ell_0\right),
    \label{fs_eq}
\end{equation}
with
\begin{equation}
    y = \frac{\left(\ell_{\rm sat} - \ell_0\right)^{1/\alpha}}{t}.
\end{equation}
Here $Y$ is a scaling function, independent of system size, and $y$ is a dimensionless scaling variable. 
In FSS analysis, one tries for collapse of data from different $L$ by varying unknown parameters which in the present case are $\ell_0$ and $\alpha$.
The values that provide optimum collapse are the accepted ones.
In Fig. \ref{doml_Lvary_fig} (d) we show such a data collapse for the considered values of $f_A$ and $T$.
The presented high quality collapse is obtained for $\ell_0 = 2.25$ and $\alpha = 2.5$. 
These numbers are consistent with those obtained from the exercise in Fig. \ref{doml_Lvary_fig} (c).
Furthermore, to be consistent with expectation in the thermodynamically large system size limit, i.e., for large $y$, we expect \cite{Fisher_barber, majumder2011_2, TP2024}
\begin{equation}
    Y(y) \sim y^{-\alpha}.
    \label{largey_eq}
\end{equation}
The data in the collapsed plot, in the large $y$ regime, are indeed consistent with this expectation.
Given that the conclusions from FSS and instantaneous exponent convergence agree quite well with each other, our estimation, henceforth, of $\alpha$ will be from the extrapolation of $\alpha_i$, to the $\ell \to \infty $ limit, by excluding the finite-size affected regions of data sets, as done in Fig. \ref{doml_Lvary_fig} (c).

Combining Eq \eqref{largey_eq}, the large $y$ behavior of $Y$, with the fact that $Y$ tends to a constant as $y \to 0$, an analytical function was constructed with the form \cite{SKD_RSC2021, skd2024, TP2024}
\begin{equation}
    Y(y) = Y_0 \left( b + \frac{y^{\theta}}{\alpha} \right)^{-\alpha/\theta}.
    \label{Y_eq}
\end{equation}
It should be noted here that $\alpha$ is already known and $Y_0$, an amplitude, can be readily estimated from observation of data in  Fig. \ref{doml_Lvary_fig} (d).
The true unknowns, $b$ and $\theta$, that should be obtained from fitting exercise, provide information on the severity of the finite-size effects.
In Fig. \ref{doml_Lvary_fig} (d), the continuous line is a fit of this function to the combined data set.

\begin{figure}
    \centering
    \includegraphics*[width=0.48\textwidth]{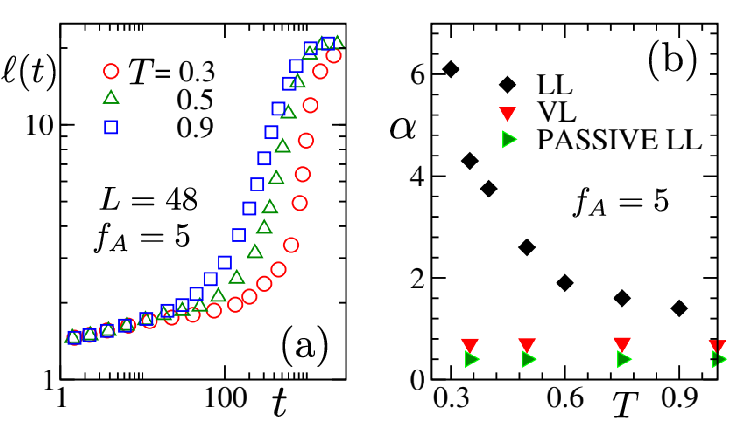}
    \caption{ (a) Plots of the average domain lengths, $\ell (t)$, versus $t$, on a log-log scale, for various temperatures, with $L=48$ and $f_A=5$. 
    (b) Growth exponents, $\alpha$, are plotted as a function of temperature.
    We have also included data for pure passive binary mixture and single component active matter, with $f_A=5$, the latter undergoing vapor-liquid transition.
    When $1/\alpha$, for the active binary mixture, is plotted versus $T$, the data exhibit a linear trend. An extrapolation cuts the abscissa at a positive value of $T$, implying the possibility of an exponential growth at low enough temperature.
    }
    \label{doml_Tvary_fig}
\end{figure}

The value obtained above for the exponent is exceptionally high, much higher than the corresponding single-component active matter undergoing vapor-liquid transition.
We have investigated whether there exists a dependence of the evolution rate on $f_A$ or $T$.
While for $f_A$ variation, the changes in $\alpha$ are unremarkable, below we discuss the case of $T$ dependence.

In Fig. \ref{doml_Tvary_fig} (a) we present direct growth data for $f_A=5$ and several values of $T$, on a log-log scale.
While it appears that with the decrease of $T$, the crossover is delayed, post-crossover period conveys a message of faster growths at lower temperatures.
In Fig. \ref{doml_Tvary_fig} (b) we show the corresponding exponents.
There is sharp rise of $\alpha$ with the decrease of $T$!
These results are compared with the single component active case, for which there exists no noticeable dependence on temperature.
For the sake of completeness, we have included data for pure passive binary mixture also.


\begin{figure}
    \centering
    \includegraphics*[width=0.48\textwidth]{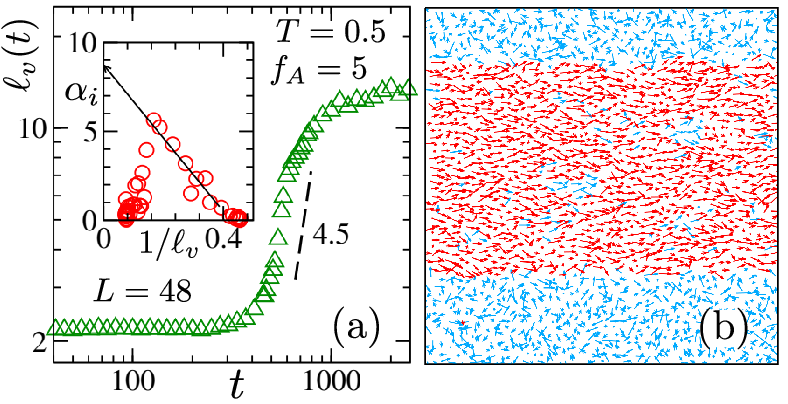}
    \caption{(a)  Plot of the length scale, $\ell_v (t)$, corresponding to the growth in the velocity field, versus $t$, on a log-log scale, for $T=0.5$, with $L=48$ and $f_A=5$.
    The inset shows the corresponding instantaneous exponent, $\alpha_i$, against $1/\ell_v$. The line with an arrowhead guides to the asymptotic limit, $\ell_v \rightarrow \infty$.
    (b) A 2D cross section showing the velocity fields of the active (red) and passive (blue) particles at a late stage of evolution. Active particles are moving coherently.
     }
    \label{vel_dom_fig}
\end{figure}

\section{Conclusion}

We have studied kinetics of phase separation in mixtures of active and passive particles. 
Owing to an interparticle potential, the model exhibits liquid-liquid phase separation even in the pure passive limit.
Activity in our model is introduced only to one of the species, via a Vicsek-like self-propulsion rule.
This enhances the rate of phase separation, keeping the basic expectations, viz., self-similar growth and finite-size scaling, valid.
The rate further increases, quite sharply, when the temperature is lowered by keeping the activity strength fixed. 
This is indicative of the possibility of even an exponentially fast evolution, at low enough temperature, confirmation of which requires simulations with very large systems.
These quantitative results are obtained via finite-size scaling and other advanced analyses.
The results have relevance in biological as well as artificial active matters, e.g., in robotic swarms \cite{swarm2013, robot2014, Poschel2018, kilbot2024}.

It is possible to achieve such remarkable phase transition dynamics if growth in a system occurs via coalescence of clusters \cite{ramanlal1985theory, carnevale1990, trizac1996, Cremer2014, Midyaprl, vadakkayil_arxiv_2026}.
If the clusters move rapidly, particularly having high degree of fractal feature, the growth rate can be very high.
Corresponding theoretical picture in fact includes the possibility of exponentially fast rate \cite{Cremer2014,vadakkayil_arxiv_2026}.
However, the structure in the present study is made of percolating or interconnected domains.
Nevertheless, branches in the structure may exhibit rapid beating type movement, or a broken up part may move ballistically, leading to the above-mentioned coalescence phenomenon in an effective sense \cite{Cremer2014,vadakkayil_arxiv_2026}.
However, it is not straightforward to identify such dynamics, particularly in three dimensional situation.

For percolating domain morphologies there are important theories accounting for hydrodynamic mechanisms. 
In one such theory, starting from a Navier-Stokes \cite{landau1987fluid, Furukawa1987, Hansenbook, Roy2013kinetics} formalism, combination of inertial dissipation and frictional dissipation was written to be proportional to $t_0S/\eta\xi$, $t_0$ providing a time scale of collective diffusion of particles having correlation over a length scale $\xi$, $S$ being a surface tension and $\eta$ the shear viscosity. 
In the case of passive matter, temperature dependence of the above quoted ratio can be readily estimated from the available knowledge of critical singularities \cite{Fisher1967}, Stokes-Einstein-Southerland (SES) relation \cite{Hansenbook}, etc. 
Given that inertial dissipation and frictional dissipation are proportional, respectively, to $v$ and $v^2$, $v~(=d\ell/dt)$ being the interface velocity, a quantitative picture of growth exponent arises naturally. 
A qualitative hydrodynamic theory exists for the Vicsek model dynamics as well \cite{toner_tu}. 
However, for active matter systems an SES relation may be violated \cite{Aras2019} and knowledge of other parameters are severely incomplete, irrespective of self-propulsion rule. 
With improved information an accurate theoretical picture for the quantitative results presented here can be obtained. 

In Fig. \ref{vel_dom_fig} (a) we show the growth of characteristic length scale corresponding to velocity field ordering.
Related instantaneous exponent is shown in the inset. 
When compared to the corresponding density field scenario, it is clear that velocity ordering occurs much faster that drives growth in the other.
In Fig. \ref{vel_dom_fig} (b) we depict the velocity ordering pictorially.
While in the domain of active particles the velocities are ordered, such a feature is absent within the passive region.
Depending upon the differences in flow in adjacent domains, the penetrating capability and thus, the possibility of merger with domains of similar species on another side of domain of unlike particles will vary.
This differences may vary with temperature, like the interfacial tension.
Detailed studies are necessary to capture this complex aspect and thus, to understand the sharp growth feature.

\section{Acknowledgement}

The authors acknowledge computation facility of National Supercomputer Mission located in JNCASR, as well as financial support from JNCASR. 
The authors also acknowledge the use of the LAMMPS \cite{LAMMPS} simulation package for molecular dynamics simulations.

\section{Data Availability}
The data will be made available upon reasonable request to the authors.


\bibliography{library}

\end{document}